\documentclass[prl,twocolumn,aps,showpacs]{revtex4}

\usepackage{graphicx}

\begin{document}

\title{Chern semimetal and Quantized Anomalous Hall Effect in
  HgCr$_2$Se$_4$}

\author{Gang Xu, Hongming Weng, Zhijun Wang, Xi
  Dai\email{daix@aphy.iphy.ac.cn}, Zhong
  Fang\email{zfang@aphy.iphy.ac.cn}}

\affiliation{Beijing National Laboratory for Condensed Matter Physics,
  and Institute of Physics, Chinese Academy of Sciences, Beijing
  100190, China;}

\date{\today}

\begin{abstract}
  In 3D momentum space, a topological phase boundary separating the
  Chern insulating layers from normal insulating layers may exist,
  where the gap must be closed, resulting in a ``Chern semimetal''
  state with topologically unavoidable band crossings at the fermi
  level. This state is a condensed-matter realization of Weyl fermions
  in (3+1) D, and should exhibit remarkable features, such as magnetic
  monopoles and fermi arcs. Here we predict, based on first-principles
  calculations, that such a novel quantum state can be realized in a
  known ferromagnetic compound HgCr$_2$Se$_4$, with a single pair of
  Weyl fermions separated in momentum space.  The quantum Hall effect
  without an external magnetic field can be achieved in its quantum-well
  structure.
\end{abstract}

 \maketitle

Under broken time reversal symmetry, the topological phases of
two-dimensional (2D) insulators can be characterized by an integer
invariant, called Chern number~\cite{Chern}, which is also known as
the TKNN number~\cite{TKNN} or the number of chiral edge
states~\cite{Halperin} in the context of the quantum Hall effect. 2D
insulators can thus be classified as normal insulators or Chern
insulators depending on whether or not the Chern number vanishes.
Since the Chern invariant is defined only for 2D insulators, it is
natural to ask what is its analog in 3D? Starting from a 2D Chern
insulating plane (say at $k_z=0$), and considering its evolution as
a function of $k_z$, generally two situations may happen.  If the
dispersion along $k_z$ is weak, such that the Chern number remains
unchanged, the system can be viewed as the simple stacking of 2D
Chern insulating layers along the $z$ direction.  Such 3D Chern
insulators are trivial generalization of the Chern number to 3D,
which is quite similar to the weak topological insulators in systems
with time reversal symmetry. However, if the dispersion along $k_z$
is strong, such that Chern number changes as the function of $k_z$,
the system will be in a non-trivial semimetal state with
``topologically unavoidable'' band crossings located at the phase
boundary separating the insulating layers in $\vec{k}$ space with
different Chern numbers~\cite{Chern-PT,Bernevig}. This is due to the
fact that the change of Chern number corresponds to a topological
phase transition, which can happen only if the gap is closed.  From
the Kohn-Luttinger theorem, we can always expect that the band
crossings appear at the fermi level at stoichiometry.

This Chern semimetal state, if found to exist, can be regarded as a
condensed matter realization of (3+1)D chiral fermions (or called
Weyl fermions) in the relativistic quantum field theory, where the
field can be described by the 2-component Weyl spinors~\cite{Weyl}
(either left- or right-handed), which are half of the Dirac spinors
and must appear in pairs.  The band-crossing points or Weyl nodes
are topological objects in the following senses. First, since no
mass is allowed in 2$\times$2 Hamiltonian, the Weyl nodes should be
locally stable and can only be removed when a pair of Weyl nodes
meet together in the $\vec{k}$ space. Second the Weyl nodes are
``topological defects'' of the gauge field associated with the
Berry's curvature in momentum space~\cite{Jungwirth,SrRuO3,Volovik}.
The gauge field around the neighborhood of the Weyl node must be
singularly enhanced, and behaves like magnetic field originating
from a magnetic monopole~\cite{SrRuO3}. The physical consequence of
such a gauge field has been discussed in the context of anomalous
Hall effect~\cite{Jungwirth, SrRuO3,AHE} observed in ferromagnetic
(FM) metals, where the Weyl nodes, if any, are always submerged by
the complicated band structures.

In this Chern semimetal state, we may also expect unusual features
like non-closed fermi surfaces (fermi arcs) on the side surfaces.
The possible fermi arcs have been recently discussed from a view
point of accidental degeneracy, and prospected for non-collinear
antiferromagnetic pyrochlore iridates~\cite{Savarasov} by the
fine-tuning of electron correlation $U$. Since the correct $U$ and
the real magnetic ordering is still unknown, we have to wait for its
material realization. In this paper, we will show that such a novel
Chern semimetal state is actually realized as the ground state of a
known FM material HgCr$_2$Se$_4$, with only a single pair of Weyl
nodes separated in the momentum space. We further find that the
long-pursuing quantized anomalous Hall effect
(QAHE)~\cite{Haldane,Onoda-QAHE,HgMnTe,Bi2Se3-QA}, i.e., the quantum
Hall effect without an external magnetic field, can be achieved in
the quantum-well structure of HgCr$_2$Se$_4$.

HgCr$_2$Se$_4$ is a FM spinel exhibiting large coupling effects
between electronic and magnetic properties~\cite{HgCrSe-rev}. It
shows novel properties like giant
magnetoresistance~\cite{HgCrSe-MR}, anomalous Hall
effect~\cite{HgCrSe-AHE}, and red shift of optical absorption
edge~\cite{HgCrSe-OP}. Its Curie temperature $T_c$ is high (around
106$\sim$120 K), and its saturated moment is 5.64
$\mu_B$/f.u.~\cite{HgCrSe-Tc,HgCrSe-M}, approaching the atomic value
expected for high-spin Cr$^{3+}$.  Its transport behavior is
different from other FM chalcogenide spinels, like CdCr$_2$Se$_4$
and CdCr$_2$S$_4$ which are clearly semiconducting. HgCr$_2$Se$_4$
exhibits semiconducting character in the paramagnetic state but
metallic in the low temperature FM
phase~\cite{HgCrSe-MR,HgCrSe-Tran-1,HgCrSe-Tran}.  The spinel
structure (space group Fd$\bar{3}$m) can be related to the
zinc-blende and diamond structures in the following way. If we treat
the Cr$_2$Se$_4$ cluster as a single pseudo-atom (called X) located
at its mass-center, then the HgX sublattice forms a zinc blende
structure. There are two HgX sublattices in each unit cell, and they
are connected by the inversion symmetry similar to the two atoms in
the diamond structure. The pseudo atom X is actually a small cube
formed by Cr and Se atoms located at the cube corners. The cubes are
connected by corner sharing the Cr atoms. As the result, each Cr
atom is octahedrally coordinated by 6 nearest Se atoms.

\begin{figure}[tbp]
\includegraphics[clip,scale=0.055]{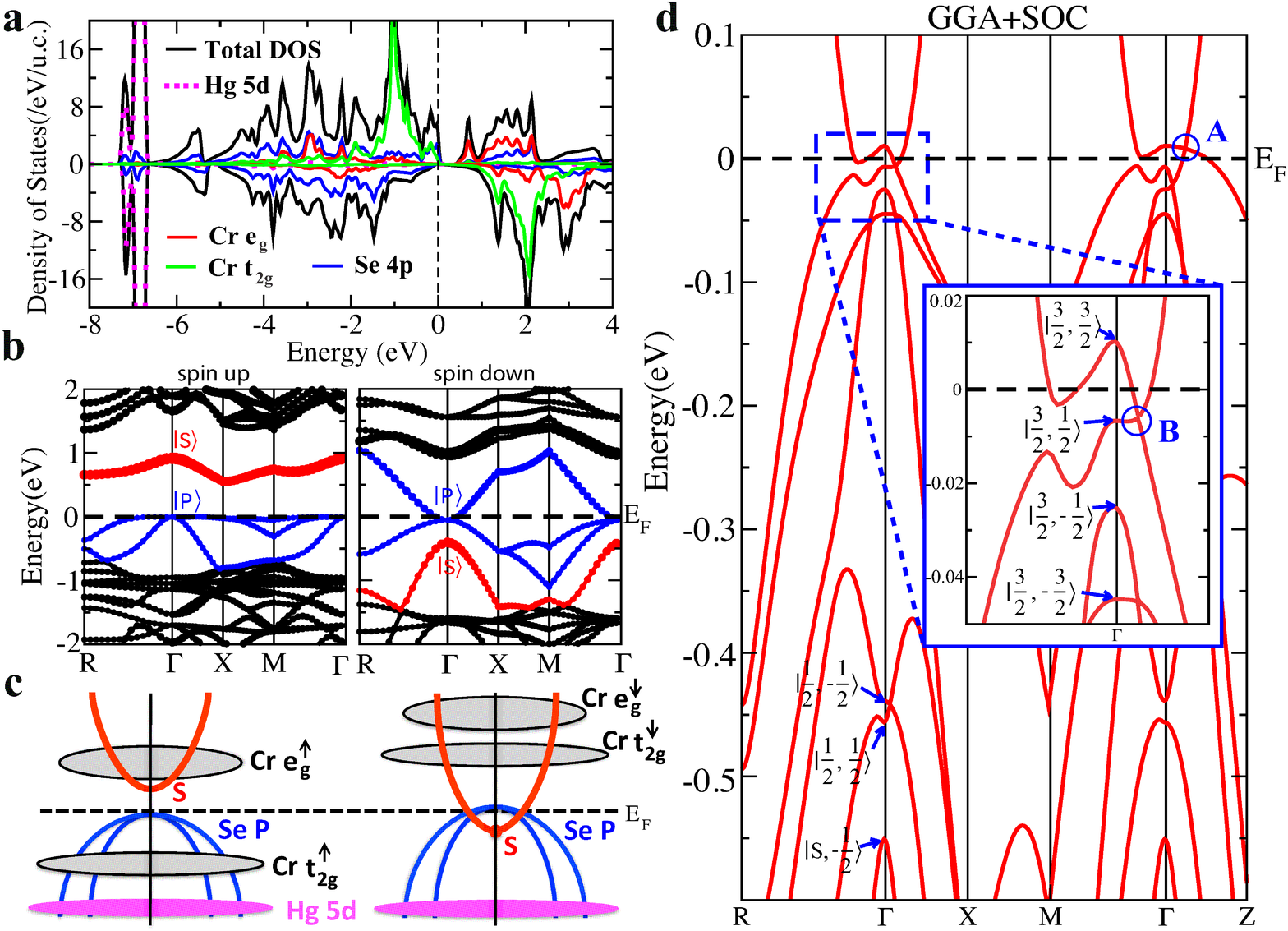}
\caption{(Color online) {\bf Electronic Structures of HgCr$_2$Se$_4$.}
  (a) The total and partial density of states (DOS); (b) The band
  structures without SOC (showing the up and down spin parts
  separately); (c) The schematic understanding for the band-inversion,
  where the $|S\rangle$ state is lower than the $|P\rangle$ states in
  the down spin channel; (d) The band structure after including SOC
  (with majority spin aligning to the (001) direction). The low energy
  states at $\Gamma$ are indicated as explained in the main text. }
\end{figure}

Our first-principles calculations~\cite{FP-method} confirm that FM
solution is considerably (2.8eV/f.u.) more stable than non-magnetic
solution, and the calculated moment (6.0 $\mu_B$/f.u) is in good
agreement with experiments~\cite{HgCrSe-Tc,HgCrSe-M}. The electronic
structures shown in Fig.1(a) and (b) suggest that the system can be
approximately characterized as a ``zero-gap half-metal'' in the case
without spin-orbit coupling (SOC). It is almost a half-metal because
of the presence of a gap in the up-spin channel just above the Fermi
level; it is nearly zero-gapped because of the band-touching around
the $\Gamma$ point just below Fermi level in the down spin channel.
The $3d$-states of Cr$^{3+}$ are strongly spin-polarized, resulting in
the configuration
$t_{2g}^{3\uparrow}e_g^{0\uparrow}t_{2g}^{0\downarrow}e_g^{0\downarrow}$.
The octahedral crystal field surrounding Cr atoms is strong and opens
up a band gap between the $t_{2g}^{3\uparrow}$ and $e_g^{0\uparrow}$
manifolds. The Se-$4p$ states (located from about -6 eV to 0 eV) are
almost fully occupied and contribute to the valence band top
dominantly. Due to the hybridization with Cr-$3d$ states, the Se-$4p$
are slightly spin-polarized but with opposite moment (about -0.08
$\mu_B$/Se). The zero-gap behavior in down spin channel is the most
important character here (Fig.1(b)), because it suggests the inverted
band structure around the $\Gamma$ point, similar to the case of HgSe
or HgTe~\cite{HgTe-1,HgTe-2}.

The four low energy states (8 after considering spin) at the
$\Gamma$ point can be identified as $|P_x\rangle$, $|P_y\rangle$,
$|P_z\rangle$, and $|S\rangle$, which are linear combinations of
atomic orbitals~\cite{AO-method}.  Considering these 4 states as
basis, we now recover the same situation as HgSe or HgTe, and the
only difference is the presence of exchange splitting in our case.
Here the band inversion (see Fig.1(c), $|S,\downarrow\rangle$ is
lower than $|P,\downarrow\rangle$) is due to the following two
factors. First, the Hg-$5d$ states are very shallow (located at
about -7.0 eV, Fig.1(a)) and its hybridization with Se-$4p$ states
will push the antibonding Se-$4p$ states higher, similar to HgSe. In
addition to that, the hybridization between unoccupied
Cr-$3d^\downarrow$ and Hg-$6s^\downarrow$ states in the down spin
channel will push the Hg-$6s^\downarrow$ state lower in energy
(Fig.1(b) and (c)). As the results, the $|S,\downarrow\rangle$ is
about 0.4eV lower than $|P,\downarrow\rangle$ states, and it is
further enhanced to be 0.55eV in the presence of SOC. We have to be
aware of the correlation effect beyond GGA, because the higher the
Cr-$3d^\downarrow$ states, the weaker the hybridization with
Hg-$6s^\downarrow$.  It has been shown that semiconducting
CdCr$_2$S$_4$ and CdCr$_2$Se$_4$ can be well described by the
LDA+$U$ calculations with effective $U$ around $3.0$
eV~\cite{CdCrS-ldau,ACrX}. We have performed the same LDA+$U$
calculations for HgCr$_2$Se$_4$ and found that the band-inversion
remains unless the $U$ is unreasonably large ($>8.0$ eV). The
experimental observations of metallic behavior at low temperature
for all kinds of samples~\cite{HgCrSe-MR,HgCrSe-Tran-1,HgCrSe-Tran}
are strong supports to our conclusion for the inverted band
structure.

\begin{figure}[tbp]
\includegraphics[clip,scale=0.05]{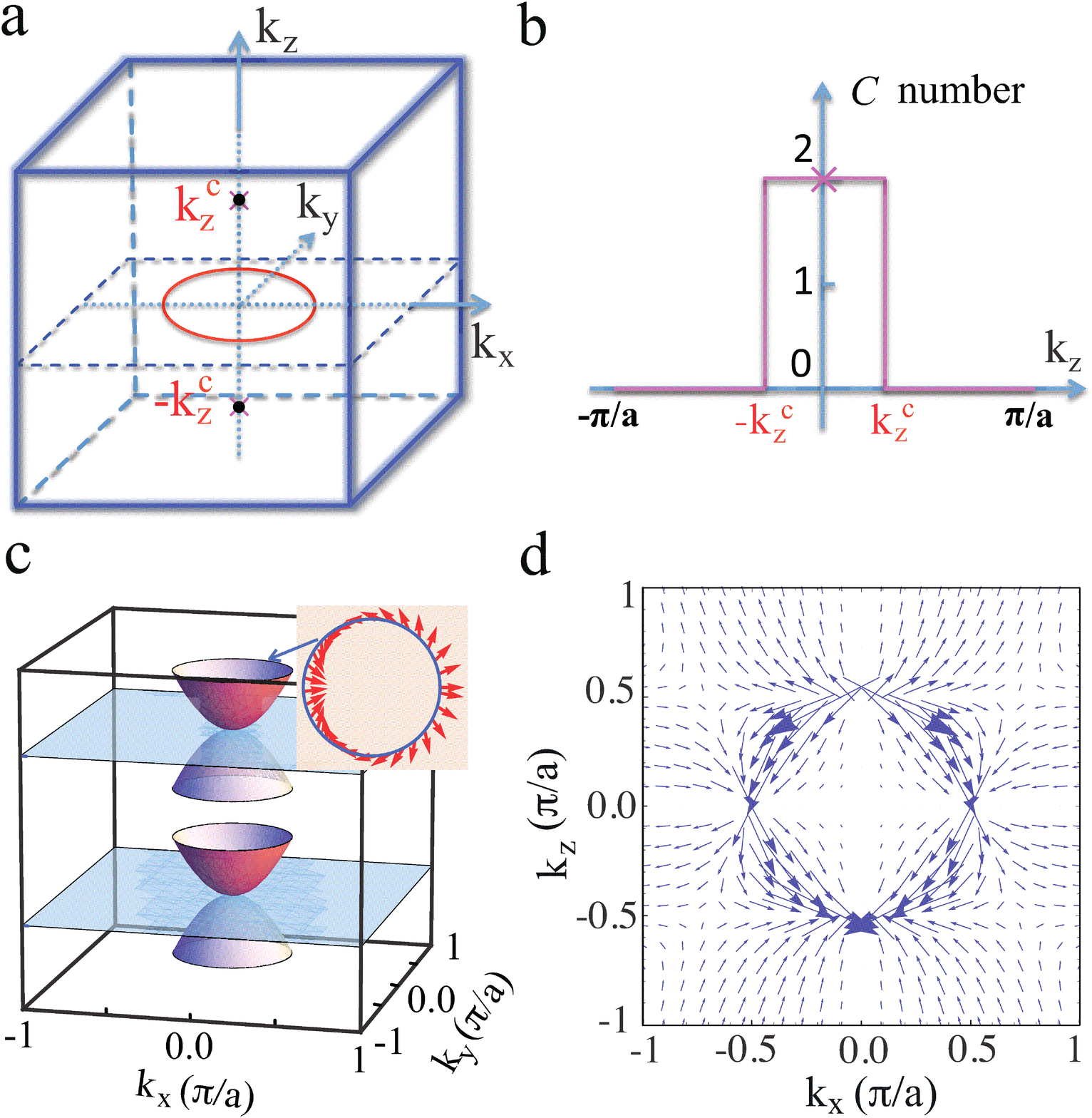}
\caption{(Color online) {\bf Weyl nodes and gauge flux in
    HgCr$_2$Se$_4$} (a) The band-crossing points in the
  $\vec{k}$-space; (b) The Chern number as function of $k_z$; (c) The
  schematic plot of the band dispersion around the Weyl nodes in the
  $k_z$=$\pm k_z^c$ plane, and the inset shows the chiral spin
  texture. (d) The gauge flux evaluated as Berry curvature in the
  ($k_x$, $k_z$) plane.}
\end{figure}

In the presence of SOC, the new low energy eigen states at $\Gamma$
are given as $|\frac{3}{2},\pm\frac{3}{2}\rangle$,
$|\frac{3}{2},\pm\frac{1}{2}\rangle$,
$|\frac{1}{2},\pm\frac{1}{2}\rangle$, and
$|S,\pm\frac{1}{2}\rangle$, which can be constructed from the
$|P\rangle$ and $|S\rangle$ states~\cite{KP}, similar to HgSe again.
Now because of the exchange splitting in our case, the eight states
at $\Gamma$ are all energetically separated, with the
$|\frac{3}{2},\frac{3}{2}\rangle$ having the highest energy, and the
$|S,-\frac{1}{2}\rangle$ being the lowest. Due to the
band-inversion, several band-crossings are observed as shown in the
band structure (Fig.1(d)). Among them, however, only two kinds of
band-crossings (called A and B) are important for the states very
close to the Fermi level. The crossing-A gives two points located at
$k_z=\pm k_z^c$ along the $\Gamma -Z$ line, and the trajectory of
crossing-B is a closed loop surrounding $\Gamma$ point in the
$k_z$=0 plane, as schematically shown in Fig.2(a).  For the 2D
planes with fixed-$k_z$ ($k_z\ne 0$ and $k_z\ne\pm k_z^c$), the band
structures are all gapped (in the sense that we can define a curved
fermi level). We can therefore evaluate its Chern number $C$ for
each $k_z$-fixed plane. We found that $C=0$ for the planes with
$k_z<-k_z^c$ or $k_z>k_z^c$, while $C=2$ for the planes with
$-k_z^c<k_z<k_z^c$ and $k_z\ne 0$.  We therefore conclude that the
crossing-A located at the phase boundary between $C=2$ and $C=0$
planes (i.e, at $k_z=\pm k_z^c$) are topologically unavoidable Weyl
nodes as addressed at the beginning.  On the other hand, however,
the crossing-B (i.e, the closed loop in $k_z=0$ plane) is just
accidental and it is due to the presence of crystal mirror symmetry
with respect to the $k_z$=0 plane. The crossing-B is not as stable
as crossing-A in the sense that it can be eliminated by changing the
crystal symmetry.

Using the 8 eigen states at $\Gamma$, we can construct an 8$\times$8
effective ${\bf k}\cdot{\bf p}$ Kane-Hamiltonian~\cite{KP}. For
qualitative understanding, however, we can downfold the 8$\times$8
Hamiltonian into a simplest 2$\times$2 model by considering the two
basis $|\frac{3}{2},\frac{3}{2}\rangle$ and $|S,-\frac{1}{2}\rangle$
which catch the band-inversion nature.
\begin{equation}
H_{eff}=\left[
\begin{array}{cc}
M &  Dk_zk_-^2  \\
 Dk_zk_+^2 & -M
\end{array}
\right]
\end{equation}
here $k_\pm=k_x\pm i k_y$, and $M=M_0-\beta k^2$ is the mass term
expanded to the second order, with parameters $M_0>0$ and $\beta>0$ to
ensure band inversion. Since the two basis have opposite parity, the
off-diagonal element has to be odd in $k$. In addition, the $k_\pm^2$
has to appear to conserve the angular moment along
$z$-direction. Therefore, to the leading order, the $k_zk_\pm^2$ is
the only possible form for the off-diagonal element. Evaluating the
eigen values $E(k)=\pm\sqrt{M^2+D^2k_z^2(k_x^2+k_y^2)^2}$, we get two
gapless solutions: one is the degenerate points along the $\Gamma -Z$
line with $k_z=\pm k_z^c=\pm\sqrt{M_0/\beta}$; the other is a circle
around the $\Gamma$ point in the $k_z=0$ plane determined from the
equation $k_x^2+k_y^2=M_0/\beta$. They are exactly the band-crossings
obtained from our first-principles calculations. Due to the presence
of $k_\pm^2$ in the off-diagonal element~\cite{Onoda-2}, it is easy to
check that Chern number $C$ equals to 2 for the planes with
$-k_z^c<k_z<k_z^c$ and $k_z\ne 0$.  The band dispersions near the Weyl
nodes at $k_z=\pm k_z^c$ plane (Fig.2(c)) are thus quadratic rather
than linear, with chiral in-plane spin texture (shown in the inset of
Fig.2(c)).  The two Weyl nodes located at $\pm k_z^c$ have opposite
chirality due to the opposite sign of mass term, and they form a
single pair of magnetic monopoles carrying gauge flux in
$\vec{k}$-space as shown in Fig.2(d). The band-crossing loop in the
$k_z=0$ plane is not topologically unavoidable, however, its existence
requires that all gauge flux in the $k_z$=0 plane (except the loop
itself) must be zero.

\begin{figure}[tbp]
\includegraphics[clip,scale=0.25]{Fig3a.eps}
\includegraphics[clip,scale=0.03]{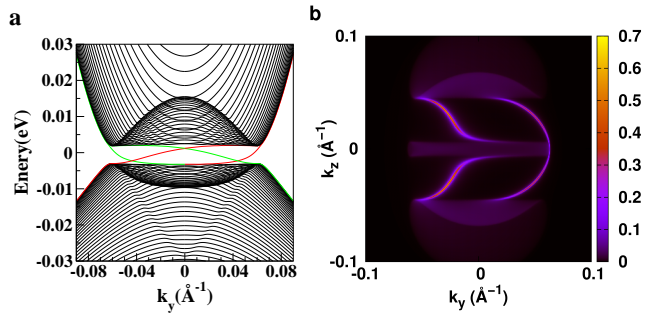}
\caption{(Color online) {\bf Edge states and Fermi arcs of
    HgCr$_2$Se$_4$~\cite{KP}.}  (a) The edge states for the plane with
  $k_z$=0.06$\pi$. A ribbon with two edges is used, and there are two
  edge states for each edge (because $C$=2). The states located at
  different edges are indicated by different colors. (b) The
  calculated Fermi arcs for the ($k_y$,$k_z$) side surface.}
\end{figure}

This Chern semi-metal state realized in HgCr$_2$Se$_4$ will lead to
novel physical consequences, which can be measured
experimentally. First, each $k_z$-fixed plane with non-zero Chern
number can be regarded as a 2D Chern insulator, and there must be
chiral edge states for such plane if an edge is created. The number of
edge states is two for the case of $C$=2 (see Fig.3(a)), or zero for
the case of $C$=0. If the chemical potential is located within the
gap, only the chiral edge states can contribute to the fermi surface,
which are isolated points for each Chern insulating plane but nothing
for the plane with $C$=0. Therefore the trajectory of such points in
the ($k_x$, $k_z$) surface or ($k_y$, $k_z$) surface form non-closed
fermi arcs, which can be measured by ARPES.  As shown in Fig.3(b), the
fermi arcs end at $k_z$=$\pm k_z^c$, and are interrupted by the
$k_z$=0 plane.  This is very much different from conventional metals,
where the fermi surfaces must be either closed or interrupted by the
Brillouin zone boundary.  The possible fermi arcs has been recently
discussed from a view point of accidental degeneracy for pyrochlore
iridates~\cite{Savarasov}.  Nevertheless, for the Chern semi-metal
state, the fermi arcs should be more stable because the band-crossings
are topologically unavoidable.

\begin{figure}[tbp]
\includegraphics[clip,scale=0.22]{Fig4a.eps}
\includegraphics[clip,scale=0.22]{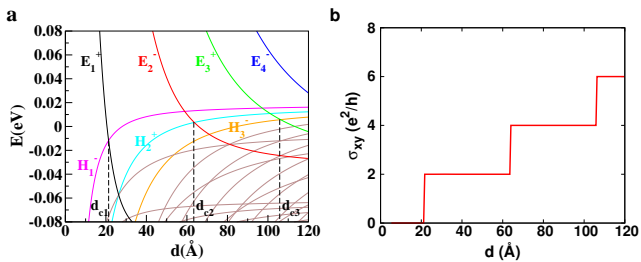}
\caption{(Color online) {\bf Quantized Anomalous Hall effect in
    HgCr$_2$Se$_4$ thin film~\cite{KP}.}  (a) The subband energy
  levels at $\Gamma$ point as function of film thickness. (b) The Hall
  conductance as function of film thickness.}
\end{figure}

The QAHE, on the other hand, is an unique physical consequence
characterizing the Chern semi-metal nature of HgCr$_2$Se$_4$, by
considering its quantum-well structure. For 2D Chern insulators, the
transverse Hall conductance should be quantized as
$\sigma_{xy}=C\frac{e^2}{h}$, where $C$ is the Chern number.  Such a
quantum Hall effect without magnetic field has been
long-pursued~\cite{Haldane,Onoda-QAHE,Bi2Se3-QA} but never achieved
experimentally. In HgCr$_2$Se$_4$, considering the $k_z$-fixed
planes, the Chern number $C$ is non-zero for limited regions of
$k_z$, and this is due to the band inversion around $\Gamma$ as
discussed above. In the quantum well structure, however, those low
energy states around $\Gamma$ should be further quantized into
subbands (labeled as $|H_n\rangle$ and $|E_n\rangle$ for hole and
electron subbands respectively), whose energy levels change as
function of film thickness. As shown in Fig.4(a), when the thickness
of the film is thin enough the band inversion in the bulk band
structure will be removed entirely by the finite size effect. With
the increment of the film thickness, finite size effect is getting
weaker and the band inversion among these subbands restores
subsequently, which leads to jumps in the Chern number or the Hall
coefficient $\sigma_{xy}$~\cite{HgMnTe}. As shown in Fig.4(b), if
the film is thinner than 21\AA (about 2 lattice constants), the
$\sigma_{xy}$ is zero; once the film thickness is larger than the
critical thickness, we find subsequent jumps of $\sigma_{xy}$ in
unit of $2e^2/h$.  In fact, the strong anomalous Hall effect has
been observed for the bulk samples of
HgCr$_2$Se$_4$~\cite{HgCrSe-AHE}. This is in sharp contrast with
pyrochlore iridates, where the anomalous Hall effect should be
vanishing due to the AF ordering.

We acknowledge the valuable discussions with Y. Ran, A. Bernevig, and
the supports from NSF of China and that from the 973 program of China
(No.2007CB925000).


\begin{references}


\bibitem{Chern} D. J. Thouless, {\it Topological Quantum Numbers in
    Nonrelativistic Physics} (World Scientific, Singapore, 1998).



\bibitem{TKNN} D. J. Thouless, M. Kohmoto, M. P. Nightingale, and
  M. den Nijs, Phys. Rev. Lett. {\bf 49}, 405 (1982).

\bibitem{Halperin} B. I. Halperin, Phys. Rev. B {\bf 25}, 2185 (1982).



\bibitem{Chern-PT} T. Thonhauser, D. Vanderbilt, Phys. Rev. Lett.,
  {\bf 74}, 235111 (2006).

\bibitem{Bernevig} T. Hughes, E. Prodan, B. A. Bernevig, arXiv:
  cond-matt/1010.4508 (2010).







\bibitem{Weyl} H. Weyl, Z. Phys., {\bf 56}, 330 (1929).



\bibitem{Jungwirth} T. Jungwirth, Q. Niu, and A. H. MacDonald,
  Phys. Rev. Lett. {\bf 88}, 207208 (2002).

\bibitem{SrRuO3} Z. Fang, et.al., SCIENCE, {\bf 302}, 92 (2003).

\bibitem{Volovik} G. E. Volovik, JETP Lett., {\bf 75}, 55 (2002).

\bibitem{AHE} N. Nagaosa, J. Sinova, S. Onoda, A. H. MacDonald,
  N. P. Ong, Rev. Mod. Phys. {\bf 82}, 1539 (2010).




\bibitem{Savarasov} X. G. Wan, A. M. Turner, A. Vishwanath,
  S. Y. Savrasov, Phys. Rev. B {\bf 83}, 205101 (2011).




\bibitem{Haldane} F. D. M. Haldane, Phys. Rev. Lett. {\bf 61}, 2015
  (1988).

\bibitem{Onoda-QAHE} M. Onoda, N. Nagaosa, Phys. Rev. Lett. {\bf 90},
  206601 (2003).

\bibitem{HgMnTe} C. X. Liu, X. L. Qi, X. Dai. Z. Fang, S. C. Zhang,
  Phys. Rev. Lett. {\bf 101}, 146802 (2008).

\bibitem{Bi2Se3-QA} R. Yu, W. Zhang, H. J. Zhang, S. C. Zhang, X. Dai,
  Z. Fang, Science, {\bf 329}, 61 (2010).



\bibitem{HgCrSe-rev} P. J. Wojtowicz, IEEE Trans. on Magn., {\bf 5},
  840 (1969).


\bibitem{HgCrSe-MR} N. I. Solin, V. V. Ustinov, S. V. Naumov,
  Phys. Solid State, {\bf 50}, 901 (2008).

\bibitem{HgCrSe-AHE} N. I. Solin, N. M. Chebotaev, Phys. Solid State,
  {\bf 39}, 754 (1997).

\bibitem{HgCrSe-OP} T. Arai, M. Wakaki, S. Onari, K. Kubdo, T. Satoh,
  T. Tsushima, J. Phs. Soc. Jpn. {\bf 34}, 68 (1973).



\bibitem{HgCrSe-Tc} P. K. Baltzer, H. W. Lehmann, M. Robbins,
  Phys. Rev. Lett. {\bf 15}, 493 (1965).

\bibitem{HgCrSe-M} P. K. Baltzer, P. J. Wojtowicz, M. Robbins,
  E. Lopatin, Phys. Rev {\bf 151}, 367 (1966).


\bibitem{HgCrSe-Tran-1} H. W. Lehmann, F. P. Emmenegger, Solid State
  Comm. {\bf 7}, 965 (1969).


\bibitem{HgCrSe-Tran} A. Selmi, A. Mauger, M. Heritier,
  J. Mag. Mag. Mat., {\bf 66}, 295 (1987).




\bibitem{FP-method} We used the WIEN2k package with the generalized
  gradient approximation (GGA). The experimental structure
  parameters~\cite{HgCrSe-M} are used, and the Brillouin zone is
  sampled with 11$\times$11$\times$11 $k$-point mesh.





\bibitem{HgTe-1} A. Delin, Phys. Rev. B {\bf 65}, 153205 (2002).

\bibitem{HgTe-2} C. Y. Moon, S. H. Wei, Phys. Rev. B {\bf 74}, 045205
  (2006);






\bibitem{AO-method} The low energy states are:
  $|P_{\alpha}\rangle\approx\frac{1}{\sqrt{8}}\sum_{i=1}^{8}|p_{\alpha}^{i}\rangle$,
  and $|S\rangle\approx 0.4\sum_{j=1}^{2}|s^{j}\rangle+0.24
  \sum_{k=1}^{4} |d_{t_{2g}}^{k}\rangle$, where $\alpha=x,y,z$, and
  $i,j,k$ runs over Se, Hg, Cr atoms in the unit cell respectively;
  $|s\rangle$, $|p_{\alpha=x,y,z}\rangle$,
  $|d_{t_{2g}=xy,yz,zx}\rangle$ are corresponding atomic orbitals of
  each atom.






\bibitem{CdCrS-ldau} C. J. Fennie, K. M. Rabe, Phys. Rev. B {\bf 72},
  214123 (2005).

\bibitem{ACrX} A. N. Yaresko, Phys. Rev. B {\bf 77}, 115106 (2008).



\bibitem{Onoda-2} M. Onoda, N. Nagaosa, J. Phys. Soc. Jpn. {\bf 71},
  19 (2002).










\bibitem{KP} Lok C. Lew Yan Voon, M. Willatzen, {\it The kp method:
    Electronic Properties of Semiconductors}, (Springer, Berlin,
  2009). See pages 57-58 for the definition of basis and 8-bands
  Kane-Hamiltonian. The parameters fitted to our first-principles
  calculations are given as $E_0$=0.174eV, $\Delta_0$=0.352eV,
  $P$=2.592eV\AA, $m_s$=0.5424$m_0$ and $m_p$=-2.966$m_0$. Two more
  parameters, describing the exchange splitting of electron and
  valence bands, are $h_s$=0.666eV and $h_p$=0.040eV. Replacing $k_x$
  by $-i\hbar\partial_x$ and using open boundary condition, we can
  diagonalize the Hamiltonian for each fixed-$k_z$, and obtain the
  edge states. If we consider the open boundary condition along $z$
  direction, using the same strategy, we can evaluate the Hall
  conductance in the quantum well structure. The QAHE is further
  confirmed by tight-binding calculations constructed from maximumly
  localized wannier functions.



\end{references}
\end{document}